\documentstyle[11pt]{article}

\begin{document}
\thispagestyle{empty}
\baselineskip=18pt

\rightline{CLNS-98/1543}
\rightline{hep-th/9802153}
\vskip 2cm

\centerline{\Large\bf BPS Mass, Dirichlet Boundary Condition,}
\centerline{\Large\bf and the Isotropic Coordinate System}
\vskip  1cm

\centerline{\large Piljin Yi\footnote{\large\tt electronic mail: 
piljin@mail.lns.cornell.edu}}
\centerline{\it F.R. Newman Laboratory of Nuclear Studies}
\centerline{\it Cornell University, Ithaca, New York 14853-5001}

\vskip 2cm

\centerline{\large\bf Abstract}
\begin{quote}
We consider test strings and test branes ending on D$p$-branes ($p\le 6$) 
and NS5-branes in the background, for a heuristic understanding of the 
dynamics. Whenever some supersymmetry is preserved, a simple BPS bound 
appears, but the central charge in question is measured by certain isotropic 
coordinate system, rather than by the actual spacetime geometry. This way, 
the ground state energy is independent of the gravitational radii of 
the solitonic background. Furthermore, a perturbation around 
the supersymmetric ground states reveals that the appropriate Dirichlet 
boundary condition is {\it dynamically} induced. We close with comments.

\end{quote}

\newpage

\section{Motivation}

In recent years, it became quite clear that fundamental strings are simply
one of many different kind of branes. In particular, branes carrying 
Ramond-Ramond antisymmetric tensor charges have emerged as  equally
important building blocks of string theories. These objects
are called D-brane, a named derived from the fact that they impose
a Dirichlet boundary condition for open fundamental strings \cite{d}. 
Such phenomena of Dirichlet boundary condition are not restricted
to the case of fundamental string ending on D-branes. Rather, D-brane
themselves can end on other kind of D-branes \cite{strominger}, and also 
D-branes may end on NS5-branes \cite{hanany}.

Physics behind open brane configurations, such as charge conservation, 
has been understood to some degree, and exploited at great length for
studying supersymmetric Yang-Mills field theories. However, there are 
more details that need be further clarified. For instance, consider the
well-known D-brane dynamics, which can be seen as induced by open strings 
stretched between them.  In the low energy limit, the primary effect is 
induced by open strings in the ground states, and the dynamics is 
known to be described by Super-Yang-Mills theories in Coulomb phase 
\cite{witten}.  One could understand this result from quantizing open 
strings as if the ambient geometry is completely trivial except a
Dirichlet boundary condition imposed strings so that it always ends
on the said D-branes 
\cite{TASI}. Nevertheless, this prescription reproduces, via a loop 
expansion, some long-range curvature effects one D-brane causes on another 
\cite{bachas}\cite{kabat}.

A possible answer to this apparent quandary is that perhaps we did quantize 
the open string in the true curved background of D-branes but did not realize 
it because we were dealing with supersymmetric configurations only. Could 
we have started with the curved background induced by D-branes, and reached 
at the same low energy D-brane dynamics? But this begs for more questions.
Supersymmetry tells us that the ground state energy of the 
stretched open string is determined by the length of the string. But which 
length? Given the curved background, do we mean the proper length? More
generally what do we by ``positions'' of D-branes? 

Our aim is to present a simple classical picture of what happens when a 
brane ends on another brane in a supersymmetric fashion. When the open string
or D-branes end on branes of sufficiently higher dimensions, we may treat the 
former as test objects in the solitonic background of the latter and study
its world-volume dynamics. A crucial fact that goes a long way clarifying
some issues, is the existence of the so-called isotropic coordinate system.
It is precisely with respect to this coordinate system that the  notion
of distances and positions must be defined, it turns out. 
We will also find that, although
the ground state of stretched branes are insensitive to the curvature
effect, the oscillator modes are not. As a byproduct, we recover the 
classical dynamics itself {\it impose} the expected Dirichlet boundary 
condition naturally on the stretched test object. We close with some
afterthoughts.

\section{Open String and Open Branes}

All the branes in string theory, including the fundamental string itself,
appear as solutions
to the low energy supergravity. The BPS ground state solution can be found by 
requiring some fraction of SUSY to be preserved. There are spinors 
associated with the unbroken SUSY whose asymptotic values satisfy simple 
algebraic conditions \cite{hanany}\cite{TASI}.
\vskip 0mm
\[
\begin{array}{lrcrrcr}
{\rm F1} \hskip 2cm 
& \Gamma^0\Gamma^1\eta_L    & = & \eta_L, & \Gamma^0\Gamma^1
\eta_R & = & - \eta_R  \\
{\rm NS5} & \Gamma^0\cdots\Gamma^5\eta_L & = & \eta_L, & 
\Gamma^0\cdots\Gamma^5\eta_R & = & \pm  
\eta_R \\
{\rm D}p & \Gamma^0\cdots\Gamma^p\eta_L & = & \eta_R. &&&
\end{array}
\]
\begin{quote}
{\bf Table 1.} Conditions on asymptotic value of spinors for unbroken SUSY
\end{quote}
where the 10-dimensional spinors satisfy the usual chirality 
condition: $\Gamma^{11}\eta_L = \eta_L, \Gamma^{11}\eta_R = - \eta_R$ for IIA 
and $\Gamma^{11}\eta_{L,R} = \eta_{L,R}$ for IIB. The sign for NS5 branes
is such that the 6-dimensional world-volume supersymmetry of NS5 branes 
is chiral $(2,0)$ for IIA and nonchiral $(1,1)$ for IIB.
When two branes intersect, or when a brane ends on another, we expect the
conditions for both branes to be satisfied simultaneously. We chose
$(\Gamma^0)^2=-1$.

The simplest supersymmetric configuration is that of a fundamental string
(F1) ending on a D-brane. From this, we can reconstruct other
configurations by duality chasing. {\bf Table 2.} summarizes the result when
the branes involved are at most of 6+1 dimensions. In this note, we will not 
consider the other cases, where the behavior of the background solutions 
are qualitatively different.

\begin{center}
\begin{tabular}{rclcrclcrcl}
    &     &     &        & F1 & on  & D6 &        &  &    & \\
 D1 & on  & NS5 &  $\leftarrow$ U $\rightarrow$       
                         & F1 &     & D5 &        &  &     & \\
 D2 &     & NS5 &        & F1 &     & D4 &        &  &     & \\
 D3 &     & NS5 &        & F1 &     & D3 & $\leftarrow$ U $\rightarrow$
                                                  & D1 &  on   & D3\\
 D4 &     & NS5 &        & F1 &     & D2 &        & D2  &    & D4   \\   
 D5 &     & NS5 &        & F1 &     & D1 &        & D3   &     & D5   \\
( D6 &    & NS5 ) &      & ( F1   & & D0 )&        & D4   &     & D6   \\
\end{tabular}
\end{center}
\begin{quote}
{\bf Table 2.} Supersymmetric configurations of brane ending on brane.
$SL(2,Z)$ U-duality of IIB theory is used to jump across the columns, 
while T-duality relates items within each of three columns.
\end{quote}

Finally,  the cases of D6 ending on NS5 or F1 ending on D0, which can be found
by a naive T-duality, cannot be realized unless there is a 
cosmological constant in the background as well. But since we will treat
open D6 and open F1 strictly as test objects, one may consider the following 
consideration applicable for those as well.

\section{Branes as Supergravity Background}

There are two distinct classes of backgrounds where
branes can end, namely those of D-branes and NS5-branes. Both are realized 
as extremal limits of black $p$-brane solutions cataloged by Horowitz and 
Strominger\cite{horowitz}. Let $X^i$ ($i=p+1,\dots, 10$) be spatial 
coordinates orthogonal to the brane and $Y^n$ ($n=1,\dots,p$) be those
along the brane. $T$ is the time coordinate.

The D$p$-brane solutions are characterized by a single harmonic function 
$H_p$ with isolated sources at $\vec{X}=\vec{X}_a$'s. For $p\le 6$, we have
\begin{equation}
H_p= 1+\sum_a \frac{k_p}{|\vec{X}-\vec{X}_a|^{7-p}},
\end{equation}
where the $k_p$'s are appropriately quantized and of dimension
of length to the power of $7-p$. In the string frame, 
the D$p$-branes has the following universal form,
\begin{equation}
G=H_p^{-1/2}(-dT^2+d\vec{Y}^2)+H_p^{1/2}d\vec{X}^2.
\end{equation}
The same harmonic function $H_p$ dictates the dilaton behavior;
\begin{equation}
e^{\phi}= {g_s}H_p^{(3-p)/4} ,
\end{equation}
with the asymptotic string coupling constant $g_s$,
and similarly determines the $(p+1)$-form field or its
dual in the Ramond-Ramond sector. See \cite{stelle} for a review.

One interesting aspect of these supersymmetric solitons is 
that the solution can be written in an isotropic manner as above. 
Note that the D-brane geometry is asymptotically flat provided $p\le 6$.
For small enough gravitational size or for large enough distances from
solitonic core, the isotropic coordinates $(T,\vec{Y},\vec{X})$ play the 
role of asymptotically flat coordinate with respect to which far away
closed strings can be quantized. Later, we will see that the same
$X^i$ and $Y^n$ coordinates appear flat as seen by the lowest 
lying open string ending on the D-brane soliton, as well, despite their
proximity.

NS5-brane solutions are even simpler and can be again written with a single
harmonic
function $\tilde H$,
\begin{equation}
\tilde{H}= 1+\sum_a \frac{\tilde{k}}{|\vec{X}-\vec{X}_a|^{2}},
\end{equation}
where
\begin{equation}
G= -dT^2+d\vec{Y}^2+\tilde{H}d\vec{X}^2,
\end{equation}
and
\begin{equation}
e^{\phi}= {g_s}\tilde{H}^{1/2}.
\end{equation}
NS5-brane is magnetically charged with respect to the NS-NS
antisymmetric tensor $B$, so that only nonvanishing components are the
$B_{ij}$'s.

\section{BPS Mass and Dirichlet Boundary Condition}

\subsection{Fundamental String on D$p$-branes}

We are interested in understanding  classical dynamics of
test objects, so we may safely ignore the fermions on the
world-sheet. One
possible form of the classical action is that of Nambu-Goto. Denoting
the induced metric on the world-sheet by $h_{\mu\nu}$, the action is
\begin{equation}
S={m_s^2}\int d\sigma^2 \sqrt{-{\rm Det}\, h} ,
\end{equation}
up to couplings to the dilaton $\phi$ and to the antisymmetric tensor $B$.
$B$ is absent in the D-brane background, while the dilaton coupling
occurs at higher order in $\alpha'=1/2\pi m_s^2$ and will be
subsequently
ignored. Calling  the spacetime coordinates $Z^I$ collectively, the 
induced metric is
\begin{equation}
h_{\mu\nu}=\partial_\mu Z^I \partial_\nu Z^J G_{IJ}.
\end{equation}
For the sake of simplicity, we will choose a static gauge where the
world-volume time $\sigma_0=\tau$ is identified with $T$.
Also we choose $\sigma_1=\sigma$ to run from 0 to 1.

Consider a pair of parallel D$p$-branes located at $\vec{X}=0$ and
$\vec{X}=\vec{L}$.
\begin{equation}
H_p= 1+\frac{k_p}{|\vec{X}|^{7-p}}+\frac{k_p}{|\vec{X}-\vec{L}|^{7-p}},
\end{equation}
Let an open string segment be stretched between
such a pair. The induced metric consist of three pieces
\begin{equation}
h= -H_p^{-1/2}d\tau^2 + H_p^{-1/2}\partial_\mu Y^n\partial_\nu Y^n
d\sigma^\nu
d\sigma^\nu +H_p^{1/2}\partial_\mu X^i\partial_\nu X^i d\sigma^\mu
d\sigma^\nu
\label{induced:string}
\end{equation}
Taking the determinant, we find
\begin{equation}
-{\rm Det}\,h= (\partial_\sigma X^i)^2+H_p^{-1}(\partial_\sigma Y^n)^2
-{\rm Det} \left( H_p^{-1/2}\partial_\mu Y^n\partial_\nu Y^n+
H_p^{1/2}\partial_\mu X^i\partial_\nu X^i \right)
\end{equation}
Note that the third term contains two factors of time derivatives $
\partial_\tau X$, $\partial_\tau Y$. This implies that there exists
the following static solution
\begin{equation}
\vec{X}=\sigma\vec{L},\quad \partial_\mu\vec{Y}={0},
\end{equation}
which corresponds to a straight BPS string segment that is located
at some constant $\vec{Y}$. The action per unit time for a static 
configuration is the energy, so we find the ground state energy to be
\begin{equation}
m_s^2L,
\end{equation}
with $L=|\vec{L}|$.
We find the BPS mass of the stretched open string is insensitive to the
gravitational size of the background. But there is a subtlety here in
that the distance that enters is not the proper distance but rather a
coordinate distance in a well-defined isotropic coordinate system.

When one considers the two D-branes themselves as dynamical objects, their
non-relativistic motion is dictated by a Yang-Mills theory \cite{witten}. 
``Positions''
are realized as the eigenvalues of the adjoint $U(2)$ Yang-Mills field,
while other heavy fields, of mass proportional to inter-brane
``distances,'' reproduces the effect of the open BPS strings stretched 
among them. A comparison with the above clearly shows that ``positions'' 
and ``distances'' in the Yang-Mills description are based on their
counterparts
in the isotropic coordinate system. The lowest lying open string appears 
quite insensitive to the actual spacetime geometry induced by D-branes.

(One may have expected to find BPS mass proportional to the proper length
of the string as measured by the string metric, which would have given
a different answer. In fact, the proper length of the string
is divergent due to a rather singular behavior of the $p$-brane
extremal horizon.)

In case of D0-branes, 
this identification between isotropic coordinates and eigenvalues of 
Yang-Mills scalar matrices is implicit in the test of M(atrix) theory 
\cite{Matrix} against 11-dimensional supergravity, where, in effect, 
loop amplitudes induced by these open BPS string modes reproduce the 
leading long-range interaction between the two D-branes 
\cite{bachas}\cite{Matrix}\cite{bbpt}.

Consider small fluctuations of the open string\footnote{A similar 
consideration  can be found in \cite{sangmin}.} around this
supersymmetric ground state.
Let $\vec{X}=\sigma\vec{L}+\vec{f}(\tau,\sigma)$ with $\vec f$
orthogonal
to $\vec{L}$, and $\vec{Y}=\vec{g}(\tau,\sigma)$. To the first
nonvanishing
order, the determinant can be expanded as follows,
\begin{eqnarray}
{\rm Det}\,h \:=\: -L^2 &+& L^2\,(\partial_\tau g ^n)^2-H_p^{-1}
(\partial_\sigma g ^n)^2\\&+& H_p L^2\,(\partial_\tau f ^i)^2-
(\partial_\sigma f ^i)^2 \\
&+&\cdots .
\end{eqnarray}
The ellipsis represents terms at least quartic in the small
fluctuations,
and $H_p$ here is actually $H_p(\vec{X}=\sigma\vec{L})$. The Lagrangian
is obtained by taking square root and expanding in small $f$ and $g$.

Note that, for fluctuations orthogonal to the background D$p$-branes, the 
combination $H_pL$ is the effective (inertial) mass density. A finite energy
motion must have a square integrable $H_p\,(\partial_\tau f ^i)^2$, and in
particular for any eigenmode of the Hamiltonian, $H_p\vec{f}^2$ must be
square-integrable. With the divergence of $H_p\sim (\Delta\sigma)^{p-7}$
near either end of the string, this immediately implies that $\vec{f}=
\vec{X}-\vec{L}$ must obey the Dirichlet boundary condition wherever it
ends on the D$p$-brane. Note that, in contrast, no such condition is imposed 
on the other fluctuations $\vec{Y}=\vec{g}$, which are parallel to the
background D$p$-brane.

The $f^i$ perturbation
represents fluctuation away from the background D$p$-brane. Thus,
the string cannot break away from the D$p$-brane, in that the end points
cannot move away from $\vec{X}=0$ or $\vec{L}$ without costing an infinite
amount of energy. One may be troubled by the fact that $f^i$ isn't by itself
the proper distance. However,
if the test string is to break away completely from D-brane, its 
boundary should be able to escape the gravitational radius of the latter. 
And this happens only if one can have finite $f^i=\delta X^i$ while 
approaching the boundary, which is impossible. 

Perhaps more to the point, 
the identification of $L$ as the string ``length'' above has shown that,
in translating to the standard prescription for D-brane conformal field 
theory, the $X^i$'s themselves should correspond to the world-sheet fields 
that must be quantized with Dirichlet boundary condition. In this sense,
we have derived the Dirichlet boundary condition from classical dynamics.

\subsection{D$(p-2)$-branes on D$p$-branes}

Other classes of allowed configurations include D$(p-2)$-brane ending
on  background D$p$-branes. From the viewpoint of D$p$-brane world-volume, 
such an endpoint behaves as magnetic monopoles and this situation 
has been studied in various contexts.

The world-volume action of D$(p-2)$-brane is the celebrated Dirac-Born-Infeld
action \cite{BI}, but again as with fundamental string we will ignore the part
irrelevant to the classical motion,
\begin{equation}
S_{p-2}=T_{p-2}\int d\sigma^{p-1} g_s\, e^{-\phi}
\sqrt{-{\rm Det}\, (h + b + m_s^{-2}{\cal F} )},
\end{equation}
where $h$ is again the induced metric and $b$ is the pull-back of the
antisymmetric tensor $B$ to the world-volume. $B$ does not appear in
the supergravity solution of D$p$-branes, so it is safe to ignore $b$
as well. $\cal F$ is the world-volume $U(1)$ gauge field strength.
The D$p$-brane tension is $T_{p}=2\pi \,(m_s/\sqrt{2\pi})^{p+1}/g_s$.

Again we choose a static gauge where $\sigma_0=\tau$ is identified with
$T$, and will consider $\vec{X}$ and $\vec{Y}$ as dynamical. The
background
remains identical to the one we considered previously, so we have a pair
of D$p$-branes located at $\vec{X}=0$ and $\vec{L}$. The induced metric
of a stretched D$(p-2)$-brane is similarly given as in 
Eq.(\ref{induced:string}),
\begin{equation}
h= -H_p^{-1/2}d\tau^2 + H_p^{-1/2}\partial_\mu Y^n\partial_\nu Y^n
d\sigma^\nu
d\sigma^\nu +H_p^{1/2}\partial_\mu X^i\partial_\nu X^i d\sigma^\mu
d\sigma^\nu .
\end{equation}
It is not difficult to see that there exists a static solution with
${\cal F}\equiv 0$, and
\begin{equation}
\vec{X}=\sigma_1\vec{L},
\end{equation}
while remaining noncompact $p-3$ directions of the D$(p-2)$-branes are
stretched parallel to the D$p$-brane. (That is, along $Y^n$ coordinates.)
Evaluating the action density on such a configuration, we find
\begin{equation}
T_{p-2}\; g_s \,e^{-\phi} \sqrt{-{\rm Det}\, h} \: d\sigma^{p-1}
\rightarrow T_{p-2}\;d\tau\wedge L d\sigma_1\wedge
V_{p-3} ,
\end{equation}
where the volume form $V_{p-3}$ is induced from the flat metric
$d\vec{Y}^2$.
The action density contains an extra factor $H_p^{(p-3)/4}$ from the
dilaton, when compared to the analog in the string
case, but this is exactly canceled by contribution from extra noncompact
$\sigma_2,\dots,\sigma_{p-2}$ directions. Thus, we again find that
the BPS mass density is given by the naive formula, except 
that the central charges are measured in terms of the isotropic coordinate 
distance rather than in terms of the proper distance.

Fluctuations around this configuration can be treated similarly as in the
case of test string. For
definiteness, let ${X}^i=\sigma_1 {L}^i+{f}^i(\sigma_\mu)$, and
$Y^n=\epsilon^n\sigma_{n+1}+g^n(\sigma_\mu)$ with $\epsilon^n=1$ for
$n\le p-3$ and $\epsilon^n=0$ for the rest. With
$L^if^i=\epsilon^ng^n=0$,
we find
\begin{eqnarray}
g_s^2\,e^{-2\phi}\,{\rm Det} (h+ m_s^{-2}{\cal F})&=& -L^2 \\
&+& L^2\left( (\partial_\tau
\vec{g})^2-(\partial_{\sigma_2}\vec{g})^2-\cdots
-(\partial_{\sigma_{p-2}}\vec{g})^2\right) -H_p^{-1}
(\partial_{\sigma_1}\vec{g})^2 \\
&+& H_pL^2\left( (\partial_\tau
\vec{f})^2-(\partial_{\sigma_2}\vec{f})^2-
\cdots -(\partial_{\sigma_{p-2}}\vec{f})^2\right) -
(\partial_{\sigma_1}\vec{f})^2 \\
&+&\cdots ,
\end{eqnarray}
where we suppressed terms of higher order in $\vec{f}$ and $\vec{g}$ as
well as those with factors of $\cal F$. The latter starts at $\sim {\cal
F}^2$. Again, a finite energy eigenmode of $\vec{f}$ must be square-integrable
against the weight $H_p(\vec{X}=\sigma_1\vec{L})$ which is severely
divergent at either end of the open D$(p-2)$ brane: The boundaries of 
the test D$(p-2)$-brane are stuck on the D$p$-branes, so that
the Dirichlet boundary condition is again dynamically imposed.

\subsection{D$p$-branes on NS5-branes}

Now we consider the final case on our list, where D$p$-branes with $p\le
5$ end  on NS5-branes. Let us put a pair of NS5-branes in the background,
as given by the harmonic function,
\begin{equation}
\tilde H=1+\frac{\tilde k}{|\vec{X}|^2}+\frac{\tilde
k}{|\vec{X}-\vec{L}\,|^2}.
\end{equation}
The induced metric of the stretched D$p$-brane is then
\begin{equation}
h= -d\tau^2 + \partial_\mu Y^n\partial_\nu Y^n d\sigma^\nu
d\sigma^\nu +\tilde H\partial_\mu X^i\partial_\nu X^i d\sigma^\mu
d\sigma^\nu ,
\end{equation}
while the pull-back of $B$ is given by
\begin{equation}
b=B_{ij}\,\partial_\mu X^i\partial_\nu X^j\,d\sigma^\mu d\sigma^\nu,
\end{equation}
where we used the fact that the $B_{ij}$'s are only nonvanishing components
of the NS-NS antisymmetric tensor.

Despite the presence of the antisymmetric tensor field, the same kind of
ground state solution is possible. Write ${X}^i=\sigma_1 {L}^i+
{f}^i(\sigma_\mu)$, and $Y^n=\epsilon^n\sigma_{n+1}+g^n(\sigma_\mu)$
with
$\epsilon^n=1$ for $n\le p-1$ and $\epsilon^n=0$, and we can easily
expand the determinant with the constraint $L^i f^i=\epsilon^ng^n=0$,
\begin{eqnarray}
g_s^2\,e^{-2\phi}\,{\rm Det} (h+b)&=& -L^2 \\
&+& L^2\left( (\partial_\tau
\vec{g})^2-(\partial_{\sigma_2}\vec{g})^2-\cdots
-(\partial_{\sigma_{p}}\vec{g})^2\right) -\tilde H^{-1}
(\partial_{\sigma_1}\vec{g})^2 \\
&+& \tilde H L^2\left( (\partial_\tau \vec{f})^2-(\partial_{\sigma_2}
\vec{f})^2-\cdots -(\partial_{\sigma_{p}}\vec{f})^2\right) -
(\partial_{\sigma_1}\vec{f})^2 \\
&+& L^2\tilde H^{-1}\left((B_{ij}L^i\partial_{\tau}f^j)^2-
(B_{ij}L^i\partial_{\sigma_2}f^j)^2-\cdots-
(B_{ij}L^i\partial_{\sigma_{p}}f^j)^2\right)\\
&+&\cdots.
\end{eqnarray}
We again recover the naive BPS mass formula for straight
open D$p$-branes from the leading term. 

In contrast to the previous examples there are new terms from $B_{ij}$. 
While this appears to modify the dynamics, for $B_{ij}$ has exactly the 
same divergence as $\tilde H$,
it is easy to see that the inner product with $\vec{L}$ reduces the 
divergence of the antisymmetric tensor field  and
that we may neglect its presence near the boundaries.
The effective inertial mass density for the $f^i$ modes diverges with
$\tilde H$ near the boundaries, as before, and 
the test D$p$-brane are stuck on the NS5-branes.

\section{Comments}

Adopting an elementary picture of test strings and test branes
ending on background supergravity solitons we 
illustrated how the two crucial properties of open strings and
open branes arise. Background independent central charge emerged
from the isotropic coordinate system. Most notably, the defining property
of Dirichlet boundary condition is found to be dynamically  induced, thanks to
a divergent effective {\it inertial} mass density of the test object.
The boundaries are too heavy to move along certain directions, in effect.

There could be many problems if
we want to elevate the present considerations to a more rigorous level. For 
instance, the solution to a low energy effective theory of type II string 
theories may not be trustworthy near the core due to a 
potentially divergent curvature. While we will not address such questions
fully, there are many cases when the potential problems can be removed, say,
by going to M-theory. 

Consider a fundamental string ending on a pair of
D6-branes. In M-theory, this corresponds to a membrane wrapping around
a nontrivial 2-cycle in a hyper-K\"ahler space with two Taub-NUT
centers. A Taub-NUT center, corresponding to a D6-brane, has only
a coordinate singularity. The 11-dimensional curvature tensor
is in fact bounded above by $\sim 1/R^2$, where $R$ is the radius of  the
compact 11th direction. We may proceed similarly as above, starting 
with the Nambu-Goto action for membrane.  The area computation can be found
in Ref.~\cite{sen}, whose result obviously agrees with ours.
In this picture, the Dirichlet boundary condition follows from the topology,
for one cannot move the string (or the spherical membrane) away
from the Taub-NUT centers without changing the topology of the membrane.

Another set of examples would be D4-brane ending on 
NS5-branes, or F1 ending on D2. In these cases, the Dirichlet boundary 
condition is naturally imposed since D4 and F1 grow out of 
the latter, which are M5 and M2 branes respectively, by developing a 
compact direction \cite{mqcd}. Also, Ref.~\cite{callan} depicts
how F1 grows out of a D$p$-brane as a solution to the Born-Infeld action,
where again the Dirichlet boundary condition would follow automatically.

The point is that the picture we try to present is simply one of many 
possible alternatives. It just happens to be the simplest
one that also can be applied uniformly across the board. Our purpose is not
in a rigorous proof, but rather in  giving the phenomena of branes 
ending on branes a more familiar look.

\vskip 5mm
\centerline{\bf Acknowledgment}
\vskip 5mm
The author thanks Sangmin Lee for helpful discussions, and The Korea 
Institute for Advanced Study for the hospitality. This work is 
supported in part by the National Science Foundation.

\end{document}